# Currículo interdisciplinar para licenciatura em ciências da natureza[1]

# Interdisciplinary curriculum for science teaching undergraduate course

Carlos Alberto dos Santos[2]
Instituto Mercosul de Estudos Avançados, Universidade Federal da Integração Latino-Americana, Brasil

Nora Valeiras
Facultad de Ciencias Exactas, Física y Naturales, Universidad Nacional de Córdoba, Argentina

## Resumo

Descreve-se neste trabalho uma proposta de currículo interdisciplinar para a formação de professores de ciências da natureza. O curso permite a obtenção de quatro diplomas: professor de ciências para o ensino fundamental (nomenclatura brasileira), professor de biologia, física e química para o ensino médio. O diploma de professor de ciências é obtido com a integralização de créditos oferecidos ao longo dos três primeiros anos do curso. Para cada ano subsequente é possível obter os diplomas de professor do ensino médio. Os componentes curriculares pertinentes às ciências da natureza são inteiramente interdisciplinares nos três primeiros anos. No quarto ano são oferecidas disciplinas específicas de biologia, física e química, para a respectiva formação de professor do ensino médio.

## Abstract

An interdisciplinary curriculum for science teaching undergraduate course will be described. The curriculum allows four degrees according the Brazilian educational legislation: science teacher for the middle school, biology, chemistry and physics teacher for the high school. The science teacher degree is obtained by accomplishing the three initial years syllabus. For each subsequent year it will be possible to obtain the other degrees. The components of the curriculum for the three initial years are radically interdisciplinary, with a pedagogical organization in such ways to prepare students for the subsequent year dedicated to a specific discipline (biology, chemistry or physics).

---

[1] Aceito para publicação na Revista Brasileira de Ensino de Física (2014).
[2] Autor correspondente: carlos.alberto@ufrgs.br

# Introdução

No atual estágio científico e tecnológico não cabe mais questionamentos quanto à importância da interdisciplinaridade. Em 14.8.2013, o artigo mais citado na *Web of Science* contendo a palavra-chave *biology* é classificado pela própria base de dados nas seguintes categorias: métodos de pesquisa bioquímica, biologia molecular, ciência computacional e cristalografia. Ou seja, trata-se de um artigo interdisciplinar, com metodologias previamente desenvolvidas por físicos e químicos e agora extensivamente aplicadas em estudos de biologia. Do mesmo modo, o artigo mais citado contendo a palavra-chave *chemistry* trata de cálculos de estrutura eletrônica, um tema que teve seus primórdios em estudos desenvolvidos no âmbito da física.

A questão que se coloca agora é em que medida é desejável e possível a materialização da interdisciplinaridade no plano pedagógico, mais especificamente no ensino das ciências da natureza. Ao longo das duas últimas décadas, a literatura tem exibido um esforço global nesse sentido. Tais iniciativas podem ser categorizadas em três níveis. Duas delas no nível universitário, destinadas sobretudo à formação de profissionais das áreas biomédicas e à formação de professores de ciências para a escola secundária. A terceira categoria refere-se às iniciativas no nível da educação básica, referentes às reformas educacionais atualmente sob análise. Embora o objetivo específico aqui refira-se à formação de professores de ciências, entendemos que nossas reflexões poderão ter repercussões positivas em propostas destinadas à formação de profissionais das áreas biomédicas, bem como em reformas da educação básica.

Um segmento qualitativa e quantitativamente relevante na área da pesquisa em ensino de ciências é aquele referente ao estudo de ideias científicas preconcebidas, ou conceitos espontâneos, ou concepção espontânea, ou *misconceptions*, como geralmente é referido na literatura em inglês [1-5]. Muito do que foi investigado nesses estudos, serviu de base para iniciativas multi e interdisciplinares, tendo como premissa que conceitos equivocadamente preconcebidos em uma área do conhecimento interferem negativamente na aprendizagem de outra área. Por exemplo, o conceito de energia equivocadamente apropriado em disciplinas de física, prejudicam a aprendizagem de sua aplicação na biologia. Como há quase consenso de que a abordagem de conceitos transversais, efetuada em contexto interdisciplinar facilita a aprendizagem significativa, no sentido ausubeliano [6-8], muitas iniciativas interdisciplinares contemporâneas buscam contornar efeitos prejudiciais de concepções espontâneas.

As primeiras tentativas de concepção e operacionalização de abordagens pedagógicas interdisciplinares no ensino das ciências da natureza circunscreviam-se a pequenos tópicos de integração entre duas ou mais áreas de conhecimento. Tais iniciativas sempre foram programadas para não interferirem na estrutura curricular, sendo portanto concebidas como acessórias à organização conceitual em andamento [9-13]. Mesmo limitadas, em termos de extensão curricular, essas iniciativas serviram de base para propostas mais ambiciosas, com interferências na estrutura curricular em diferentes graus de extensão e profundidade [14-17]. O resultado de tudo isso é que supostamente estamos preparados para uma intervenção mais ambiciosa na estrutura curricular de cursos de formação de professores de ciências, com vistas a uma abordagem pedagógica radicalmente interdisciplinar.

Os avanços recentes, sobretudo da nanociência, têm impactado todas as áreas do conhecimento, mas é na biologia que a repercussão tem se dado com mais intensidade. Atualmente os trabalhos de biologia mais citados na literatura científica internacional são aqueles que incorporam conceitos e técnicas advindos da física e da química. Não é por acaso que entre os dez artigos mais citados em agosto de 2013 na *Web of Science* com a palavra-chave *biology*, nove estão relacionados com conceitos e técnicas da física e da nanotecnologia. Esse fato tem levado pesquisadores, professores e gestores educacionais a um novo posicionamento frente às estruturas curriculares dos cursos de formação universitária de profissionais das ciências da vida. Espera-se para o futuro próximo, que os biólogos incorporem a estrutura conceitual e o arsenal experimental da física e da química de modo que tenham condições de desenvolver seus estudos autonomamente. A expectativa é da emergência de um profissional interdisciplinar, de modo a diminuir a dependência de uma equipe multidisciplinar na implementação de projetos mais simples, deixando para aqueles mais complexos a inevitável constituição de equipes multidisciplinares. Apenas para citar um exemplo bastante simples: o biólogo do futuro explicará a fotossíntese sabendo exatamente o que significa a equação de Einstein para o efeito fotoelétrico, o fenômeno físico que inicia a transformação da energia solar em energia química neste processo importante para a vida em nosso planeta.

## Iniciativas inspiradoras

Para enfrentar os novos paradigmas das atividades científicas e tecnológicas resultantes da revolução nanotecnológica e da interdisciplinaridade, órgãos governamentais e associações científicas em todos os continentes têm alertado para a necessidade de reformas curriculares na educação básica e no ensino superior [18-24]. Muitos pesquisadores alinhados com essas demandas, implementaram estudos relevantes para o estabelecimento de referenciais teóricos e metodológicos indispensáveis para a operacionalização das reformas pretendidas. Como era de se esperar, a maioria desses estudos trata de novas abordagens pedagógicas no ensino de biologia na escola secundária [25] e em universidades [11, 26-33]. Poucos estudos dedicam-se a propostas curriculares interdisciplinares completas, quer seja para a formação de pesquisadores [14, 34] ou de professores do ensino básico [10, 35-37].

Embora muitos estudos tenham enfatizado a importância dos conceitos de energia e matéria para o processo de transversalidade entre as ciências da natureza [38-43], ao nosso conhecimento não há qualquer proposta curricular baseada nesses conceitos. Algumas das propostas ditas interdisciplinares, são na verdade propostas multidisciplinares construídas a partir de contextos onde são inseridas as várias disciplinas. A proposta que apresentaremos a seguir é predominantemente construída a partir dos conceitos de energia e matéria. À medida que a estrutura conceitual vai evoluindo, produtos tecnológicos resultantes das manipulações dos conceitos-chaves serão abordados. A estrutura curricular atende, portanto, aos princípios CTS (Ciência, Tecnologia e Sociedade). Por exemplo, quando os conceitos de energia e matéria são tratados no contexto da termodinâmica, todos os produtos da primeira revolução industrial serão considerados, e simultaneamente os alunos terão oportunidade de participar de discussões sobre aspectos políticos, sociais e econômicos daquele período histórico.

## Cenário atual da formação de professores de ciências

No Brasil, a formação de professores de ciências para o ensino fundamental (faixa etária até 14 anos) pode ser feita nos cursos de pedagogia das faculdades de educação e nos cursos de licenciaturas em biologia, física e química ministrados nos respectivos departamentos ou institutos universitários. Inúmeras razões históricas e culturais determinam o evidente fato de que os currículos das licenciaturas das principais universidades brasileiras não correspondem aos parâmetros curriculares definidos pelo Ministério de Educação (MEC)[44]. Não cabe aqui discutir essas razões, nem apresentar um estudo exaustivo dessa falta de sintonia entre programas de formação e marco regulatório do MEC, até porque os currículos apresentam estruturas muito similares. Há pequenas variações nas terminologias e cargas horárias das disciplinas, um determinado currículo pode ter alguma disciplina não existente nos outros, mas a estrutura básica obedece um padrão médio nacional. Portanto, poderemos desenhar um cenário satisfatório se analisarmos com algum detalhe uma universidade reconhecidamente de primeira linha, como a Universidade Federal do Rio Grande do Sul (UFRGS)[45]. Para efeito de comparação entre as diferentes licenciaturas da UFRGS, os componentes curriculares obrigatórios foram divididos nas categorias abaixo relacionadas (as respectivas cargas horárias estão expostas na Tabela 1):

1. Educação (avaliação, educação especial, filosofia, história, psicologia e sociologia da educação, pedagogia, teorias de aprendizagem, e disciplinas similares);
2. Legislação e gestão;

3. Educação em ciências e matemática (não trata de conceitos específicos de biologia, física e química, trata apenas de aspectos metodológicos);
4. Biologia (disciplinas específicas, similares àquelas ministradas nos bacharelados, incluindo bioestatística e biofísica, bioquímica);
5. Física (disciplinas específicas, similares àquelas ministradas nos bacherelados);
6. Matemática (disciplinas específicas, similares àquelas ministradas nos bacherelados);
7. Química (disciplinas específicas, similares àquelas ministradas nos bacherelados);
8. TIC & Informática (Tecnologias de informação e comunicação, informática na educação, métodos computacionais).

Tabela 1 – Carga horária das diferentes categorias de componentes curriculares para alguns cursos de licenciatura da UFRGS.

|  | Pedagogia | Biologia | Física | Química |
| --- | --- | --- | --- | --- |
| Educação | 1.800 | 285 | 240 | 210 |
| Legislação | 210 | 60 | 60 | 120 |
| Edu. C&M | 240 | --- | 150 | 90 |
| Biologia | --- | 1.890 | --- | --- |
| Física | --- | 60 | 1.695 | 270 |
| Matemática | --- | --- | 330 | 240 |
| Química | --- | 90 | 60 | 1.350 |
| TIC & Info | 45 | --- | 60 | --- |

A Tabela 1 exibe uma grande assimetria na formação de professores de ciências para o ensino fundamental. De um lado, os licenciados em pedagogia não têm qualquer formação científica, uma vez que os conceitos de biologia, física e química, mesmo aqueles mais elementares não estão presentes em qualquer componente curricular. Por outro lado, o caso mais grave talvez seja dos licenciados em biologia que têm pouca formação em física e química e nenhuma formação em matemática. Tais lacunas são prejudiciais em qualquer currículo do ensino fundamental, por mais tradicional que seja. É impossível que professores de ciências naturais formados com esses currículos atendam às recomendações curriculares do MEC para o ensino fundamental, que se supõe adequado ao ingresso das novas gerações na revolucionária nanotecnologia.

Os números apresentados na Tabela 1 sugerem que os currículos são elaborados à semelhança dos respectivos bacharelados. Ou seja, são razoavelmente consistentes com o ensino médio tradicional, mas têm pouca aderência pedagógica com o ensino fundamental e com o que se imagina deverá ser o ensino médio do século XXI.

A situação não é muito diferente nos outros países da América Latina [46]. Por exemplo, a formação docente na Argentina é realizada em dois sistemas, através de políticas e ações promovidas pelo Ministério de Educação, Ciência e Tecnologia, e executadas pela Secretaria de Educação e a Secretaria de Políticas Universitárias. Os professores de ciências, formados nos Institutos Superiores de Formação Docente (ISFD) ou nas universidades, podem exercer a docência no nível primário, cuja faixa etária vai dos 6 aos 12 anos. Os professores do secundário também podem ser formados pelos ISFDs e pelas universidades. A diferença é que aqueles formados na universidade têm um currículo acoplado às "licenciaturas" (similar aos bacharelados no Brasil) no que tange às disciplinas científicas, e têm uma série de disciplinas didático-pedagógicas, específicas para a carreira de professor.

## Objetivos

Face a este cenário, propõe-se um curso de formação de professores para a educação básica com estrutura curricular interdisciplinar consistente com demandas e orientações expostas em documentos de órgãos educacionais de vários países latino-americanos. Como se verá mais adiante, a estrutura curricular tem as seguintes características:

1. Três anos iniciais absolutamente interdisciplinares em termos de conceitos científicos;
2. Em cada semestre, disciplinas de educação são oferecidas para a devida contextualização pedagógica das matérias científicas em pauta;
3. Seminários especiais com professores de economia, filosofia, geografia, história e sociologia estão previstos em cada semestre para as respectivas contextualizações das matérias científicas;
4. Com os créditos obtidos nos três anos iniciais, e a realização de estágio curricular, o aluno pode obter o diploma de professor de ciência para o ensino fundamental;
5. Para obtenção do diploma de professor do ensino médio, o aluno cursará um ano de disciplinas específicas (biologia, física ou química). Assim, ele poderá obter o diploma de professor de biologia no quarto ano, o de professor física no quinto ano e o de professor de química no sexto ano. Ou seja, o curso permite a obtenção de quatro diplomas.

Entendemos que a implementação da estrutura acima será consideravelmente facilitada se a integração curricular se der por meio de conceitos-chaves. Não é difícil mostrar que todas as questões da biologia, física e química podem ser abordadas a partir dos conceitos de energia e matéria. Um modo interessante de mostrar isso é a partir de contextos apropriados. Como ilustração, consideremos o contexto de uma usina hidroelétrica. A Figura 1 mostra parte de uma possível estrutura temática. Tendo energia e matéria como conceitos-chaves, é possível, a partir do exame da criação de um lago artificial e do aproveitamento tecnológico de uma queda d'água, discutir conceitos tradicionalmente pertencentes às disciplinas de biologia (ecologia, zoologia, botânica, genética, bioquímica), física (mecânica, termodinâmica, eletromagnetismo) e química (físico-química, química inorgânica, química orgânica). Mostraremos na sequência que é possível rearranjar a estrutura curricular convencional para coloca-la numa perspectiva interdisciplinar, de modo que possa ser usada em diferentes contextos. Não custa reforçar a ideia de que a perspectiva interdisciplinar é constituída simplesmente com o rearranjo da estrutura curricular convencional, sem qualquer perda de conceitos básicos essenciais.

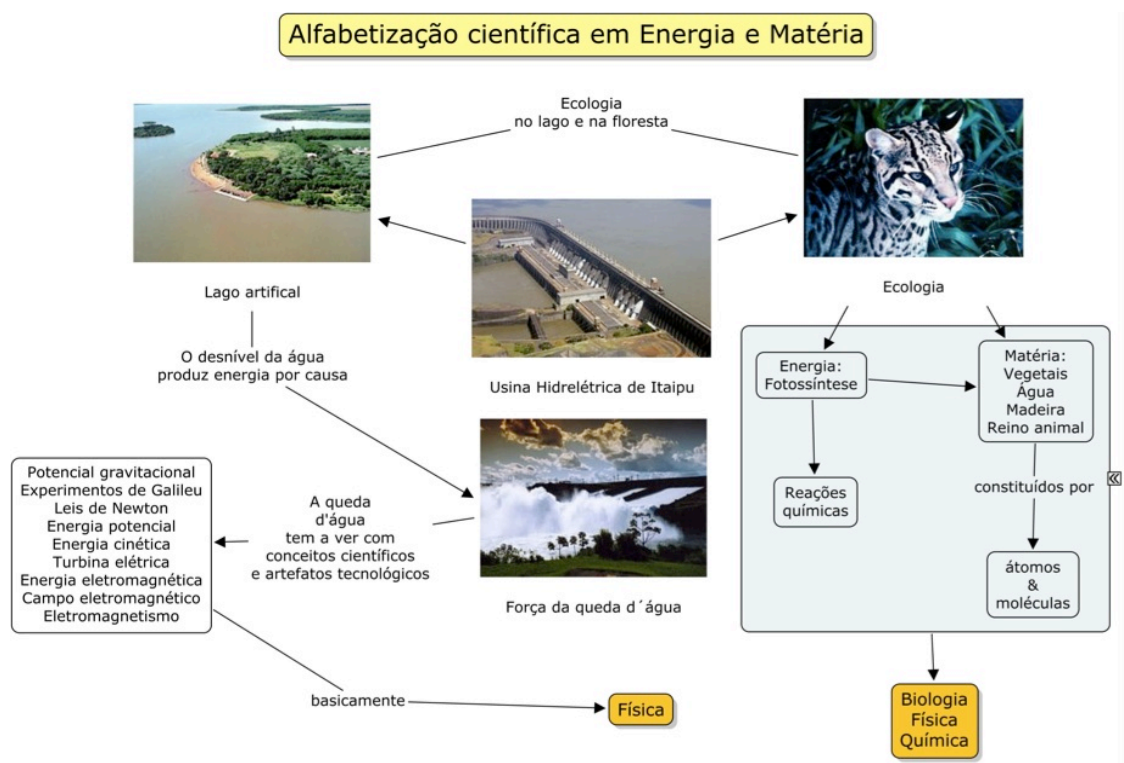

Figura 1 – Mapa conceitual a partir de um contexto definido por uma hidroelétrica. A estrutura curricular é do tipo CTS.

Trechos importantes extraídos das orientações curriculares para o ensino médio no Brasil, são reproduzidos a seguir para ilustrar o nível de correlação entre a presente proposta e as recomendações das autoridades educacionais dos países do Mercosul e de vários outros países da América Latina.

> Relacionar conceitos da Biologia com os de outras ciências, como os conhecimentos físicos e químicos, para entender processos como os referentes à origem e à evolução da vida e do universo ou o fluxo da energia nos sistemas biológicos [ref. 47, p.46].

Essa recomendação é inteiramente atendida nos três primeiros anos do curso, a etapa de formação do professor de ciências para o ensino fundamental. No quarto ano, quando o aluno opta por uma formação para o ensino médio (biologia, física ou química) ele tem contato com abordagens disciplinares específicas. O primeiro contato que o aluno tem com as disciplinas científicas é por meio de um componente curricular intitulado **Alfabetização Científica em Energia e Matéria**. Como o nome sugere, trata-se de uma abordagem em nível de popularização científica, onde praticamente todos os conceitos a serem tratados nas diversas disciplinas do curso, são inicialmente apresentados a partir de discussões de avanços tecnológicos contemporâneos da biologia, da física e da química.

O primeiro componente curricular dedicado à formalização de conceitos científicos é **Energia Gravitacional**. Parte considerável da ementa desse componente corresponde à primeira disciplina de física usualmente oferecida, com diferentes níveis de aprofundamento, nas tradicionais licenciaturas de biologia, física e química. Na presente proposta a disciplina foi organizada com uma perspectiva interdisciplinar, de modo que ela incorpora abordagens referentes à origem e à evolução da vida e do universo ou o fluxo da energia nos sistemas biológicos, como a formação da Terra e o papel da gravidade na existência de nossa atmosfera, e o efeito da gravidade na estrutura e comportamento dos seres vivos.

De modo análogo, o componente curricular correspondente à primeira disciplina sobre eletricidade e magnetismo nos cursos tradicionais, é aqui denominado **Energia Eletromagnética**, e tem, além do conteúdo usual, tópicos sobre o espectro eletromagnético e os seres vivos, interação das radiações ionizantes com materiais biológicos, sistemas de orientação e migração de diversas espécies, fototropismo, fotoactismo e biomagnetismo.

No componente curricular, **Energia nas Reações Químicas e Celulares**, conceitos de biologia, física e química são interdisciplinarmente utilizados para entender processos como os referentes à origem e à evolução da vida e do universo ou o fluxo da energia nos sistemas biológicos. Por exemplo, o tratamento da fotossíntese, evento essencial para o surgimento da vida, inicia com uma abordagem do efeito fotoelétrico e a importância da teoria quântica para a compreensão de reações fotoquímicas envolvidas na fotossíntese.

> Construir uma visão sistematizada dos diversos tipos de interação e das diferentes naturezas de fenômenos da física para poder fazer uso desse conhecimento de forma integrada e articulada. Por exemplo, reconhecer que as forças elástica, viscosa, peso, atrito, elétrica, magnética etc. têm origem em uma das quatro interações fundamentais: gravitacional, eletromagnética, nuclear forte e nuclear fraca [ref. 47, p.85].

Essa questão é atendida, no presente projeto, quando os estudantes de Biologia, Física e Química estudam as mesmas disciplinas: **Energia Gravitacional**, **Energia nas Reações Químicas e Celulares**, **Termodinâmica**, **Átomos e Moléculas**, **Energia Eletromagnética**, entre outras.

> Alguns conceitos gerais nas ciências, como os de unidades e de escalas, ou de transformação e de conservação, presentes de diferentes formas na Matemática, na Biologia, na Física e na Química, seriam muito mais facilmente compreendidos e generalizados, se fossem objeto de um tratamento de caráter unificado feito de comum acordo pelos professores da área. (…) são diferentes as conotações destes conceitos nas distintas disciplinas, mas uma interpretação unificada em uma tradução interdisciplinar enriqueceria a compreensão de cada uma delas [ref. 47, p.19].

Nessa nova compreensão do ensino médio e da educação básica, a organização do aprendizado não seria conduzida de forma solitária pelo professor de cada disciplina, pois as escolhas pedagógicas feitas numa disciplina não seriam independentes do tratamento dado às demais, uma vez que é uma ação de cunho interdisciplinar que articula o trabalho das disciplinas, no sentido de promover competências [ref. 47, p.10].

Essa articulação interdisciplinar (…) é uma dívida antiga que se tem com o aluno. Uma parcela dessa dívida poderia ser paga com a apresentação de uma linguagem e da nomenclatura realmente comuns entre várias das disciplinas. (…) quando na Biologia se fala em energia da célula, na Química (…) em energia da reação e na Física em energia da partícula, não basta que tenham a mesma grafia ou as mesmas unidades de medida. São tratados em contextos tão distintos os três temas, que o aluno não pode ser deixado solitário no esforço de ligar as "coisas diferentes" designadas pela mesma palavra. O problema da escola é que, a despeito de estarem estas três energias relacionadas, nem mesmo os professores que usam esses termos estão à vontade para interpretar seu significado em outra disciplina além da sua [ref. 47, p.19].

> Isso implica, de certa forma, um conhecimento de cada uma das disciplinas também pelos professores das demais, pelo menos no nível do ensino médio, o que resulta em uma nova cultura escolar, mais verdadeira, pois se um conhecimento em nível médio de todas as disciplinas é o que se deseja para o aluno, seria pelo menos razoável promover esse conhecimento na escola em seu conjunto, especialmente entre os professores [ref. 47, p.34].

Na formação interdisciplinar da presente proposta essas questões são inteiramente superadas, uma vez que o professor tem proficiência em biologia, física e química, obrigatoriamente no nível do ensino fundamental (três anos iniciais da licenciatura), e opcionalmente no nível do ensino médio, se obtiver os diplomas de Professor de Biologia, de Física e de Química. Ou seja, com seis anos de escolaridade universitária, o professor terá formação interdisciplinar para toda a educação básica.

> É esse contexto que dá efetiva unidade a linguagens e conceitos comuns às várias disciplinas, seja a energia da célula, na Biologia, da reação, na Química, do movimento, na Física, seja o impacto ambiental das fontes de energia, em Geografia, a relação entre as energias disponíveis e as formas de produção, na História [ref. 47, p.34].

Nesse sentido, cabe destacar os seminários especiais, a título de disciplinas eletivas ou atividades complementares da presente proposta. Por exemplo, no semestre em que se ministra termodinâmica, professores de geografia, história, economia, entre outros, ministrarão seminários sobre a primeira revolução industrial.

O documento Formación Docente Continua Inicial para la Educación Media y Tercer Ciclo – Professorado de Ciencias Básicas, do Ministerio de Educación y Cultura do Paraguai, estabelece, entre outros, os seguintes princípios[48]:

> Ademais, en el Plan específico del Bachillerato Científico con énfasis en Ciencias Básicas, a partir del Segundo Curso, se desarrollan algunas disciplinas específicas o ramas de las Ciencias Básicas ya abordadas al interior del área general, pero en este caso, con mayor profundización, con enfoque disciplinar: Física, Química, Biología, Geología y Educación Ambiental y Salud. Por otro lado, algunos proyectos educativos del Plan Optativo son disciplinas o módulos relacionados con el campo de estudios de las Ciencias Básicas (p.13).

Essa estrutura corresponde ao nosso ciclo interdisciplinar de 3 anos, durante os quais formamos professores de ciências para o ensino fundamental. A propósito, cabe aprofundar a discussão da interdisciplinaridade frente às dúvidas que eventualmente suscita, como essas expostas no documento *La Formación Docente en el Peru*[49]:

> En los medios pedagógicos peruanos ligados al Ministerio de Educación, en las Facultades de Educación y en los sindicatos magisteriales existe una concepción que sostiene que el docente principalmente debe estar formado en el *cómo enseñar* y subsidiariamente en el *qué enseñar*. Este punto de vista ha orientado progresivamente la estructuración de los currículos de formación magisterial durante los últimos 25 años, lo que explica que en los currículos oficiales aplicados a nivel nacional por la Dirección Nacional de Formación y Capacitación Docente (DINFOCAD), a través del denominado Plan Piloto de Formación Docente, hayan desaparecido totalmente las disciplinas científicas, tecnológicas y humanísticas para dar paso a carteles de contenidos, en nuestra opinión, superficiales, lógicamente desarticulados y desactualizados, a los que los funcionarios del MED denominan áreas de trabajo interdisciplinario, ignorando así el hecho del que la interdisciplinariedad no se opone a la especialización y profundización disciplinaria porque es una estrategia para integrar en el trabajo científico a equipos de especialistas en ámbitos diversificados y no para obviar la profundidad, riqueza y potencia explicativa de los conocimientos especializados (p.5).

Essa problemática foi considerada durante a elaboração do presente projeto. As abordagens interdisciplinares no curso constituem uma estratégia para integrar saberes disciplinares, na forma mais aprofundada possível. Por exemplo, ao abordar o espectro eletromagnético, os alunos trabalham inicialmente com um professor de física para apropriação dos conceitos básicos, sobretudo aqueles referentes à interação da radiação com a matéria. Mas, a abordagem do efeito das radiações ionizantes sobre os organismos vivos é feita por um professor de biologia, cuja expertise nessa área é geralmente superior à dos professores de física ou química.

No que se refere ao uso das novas tecnologias de informação e comunicação (TIC), vale destacar o documento do MEC/Paraguai já mencionado[48]:

> A fin de lograr una formación integral, pertinente, relevante, suficiente y coherente respecto a las necesidades del diseño curricular de la Educación Media y Tercer ciclo, la realidad cultural paraguaya y ante los retos de la sociedad actual y de los avances científicos y tecnológicos, el presente plan curricular toma también en consideración el abordaje del bilingüismo, los temas transversales, el desarrollo de los proyectos educativos, el uso de las tecnologías tradicionales y de las nuevas Tecnologías de la Información y la Comunicación (TICs)(p.14).

> El uso de las **tecnologías** tradicionales, y especialmente las nuevas, conocidas como Tecnologías de Información y la Comunicación (TICs**)**, deben ser promovidas en todos los momentos didácticos y en todas las actividades pedagógicas, a fin de familiarizar al/a estudiante docente con la búsqueda, obtención y procesamiento de las informaciones disponibles en las redes de comunicación, a menudo muy accesibles y beneficiosas en costo, rapidez y actualidad (p.16).
>
> Se deberá promover y promocionar el uso de la computadora, el acceso a las páginas Web, la comunicación mediante las redes informática por Internet y otros avances tecnológicos que aparecen en forma continua en este campo, pues en la era del conocimiento, pues en los contextos educativos contemporáneos la información actualizada y apropiada, preferentemente podrá obtenerse por la vía de las TICs (p.16).

O presente projeto atende inteiramente esses pressupostos, que são também defendidos nos documentos *La Formación Docente en la Republica Argentina*[50] e *Plan de Formación de Maestros 2005*[51], do Uruguai e pelas orientações curriculares para o ensino médio, do MEC[44]. A estrutura curricular aqui proposta tem três disciplinas sobre uso de informática no ensino de ciências. O primeiro é uma disciplina instrumental, sobre as ferramentas para edição de textos, planilhas, apresentações e sobre os recursos disponíveis na Internet. A segunda disciplina trata de possibilidades do uso da informática no ensino de ciências no nível fundamental. Finalmente, o terceiro componente trata do uso da informática no ensino de biologia, física ou química. Essa última disciplina é oferecida aos alunos que optarem pela respectiva carreira no nível do ensino médio.

Sobre atividades experimentais, tomemos como referência o documento do MEC/Paraguai[48], que tem premissas muito similares aos documentos dos outros países:

> La aplicación de los procesos del método científico podrá hacerse tanto en el desarrollo de las **experimentaciones** de clase de corta duración, realizadas en laboratorios reales o simulados, como en el de una **investigación de rigor científico - metodológico**. Las experiencias de laboratorio de ciencias buscarán el redescubrimiento de procesos naturales como en el descubrimiento de nuevos conocimientos, de acuerdo a la realidad y al context (p.17).
>
> En el contexto de la experimentación conciben las **salidas de campo** y las excursiones de estudios, muy útiles en el abordaje de la enseñanza de Biología, Geología, Ecología y

> Educación Ambiental, Educación para la Salud, e incluso en la enseñanza de la Física y la Química (p.17).

Essa é uma questão que merece atenção especial, e que na presente proposta tem um tratamento inovador, com o requisito inegociável de que as atividades envolvam circunstâncias de aprendizagem significativa. É muito extensa a literatura sobre a importância de atividades experimentais no ensino de ciências [52], sobre dificuldades para o uso de atividades experimentais [53], sobre diferentes metodologias de uso de laboratório no ensino de ciências [54-57], sobre alguns experimentos com materiais de baixo custo para mecânica [58], termodinâmica [59-61], ótica [62], magnetismo [63], entre outros.

Consistente com a ideia geral expressa nessa literatura, as atividades experimentais na presente proposta são encaradas como mecanismos importantes na apropriação conceitual, do mesmo modo como são encaradas pelos cientistas, e não como simples etapa instrumental do processo ensino-aprendizagem. Assim, o projeto pedagógico não prevê componentes curriculares estritamente experimentais. Os experimentos serão realizados em contexto pedagógico laboratorial, mas podem ser realizados em sala de aula ou nos laboratórios. O importante é que seja integrado ao desenvolvimento conceitual que se dá em sala de aula, sempre que a abordagem conceitual exigir a realização de uma verificação experimental. Esse é um procedimento metodológico similar à atividade científica.

Finalmente, a proposta instrumentaliza os futuros professores para uso extensivo de recursos computacionais, quer seja na realização de experimentos controlados por computador, ou na simulação de fenômenos físicos[64, 65]. Nas duas disciplinas sobre informática na educação, são abordados recursos para a elaboração e uso de hipertextos [66] e outros tipos de objetos de aprendizagem [67].

## Nossa proposta curricular

A estrutura curricular é sintética e graficamente representada na Figura 2. A denominação **Humanas**, engloba disciplinas de educação, filosofia, história, geografia, economia, entre outras, de acordo com o projeto pedagógico institucional e preferencialmente com uma perspectiva interdisciplinar, conforme exemplos apresentados mais adiante. Sob a denominação genérica **Atividades Complementares**, podem ser programados seminários, atividades de iniciação científica, participação em eventos acadêmicos, oferta de disciplinas eletivas, entre outras atividades de interesse da instituição. Vamos discutir um pouco mais detalhadamente as várias disciplinas do curso.

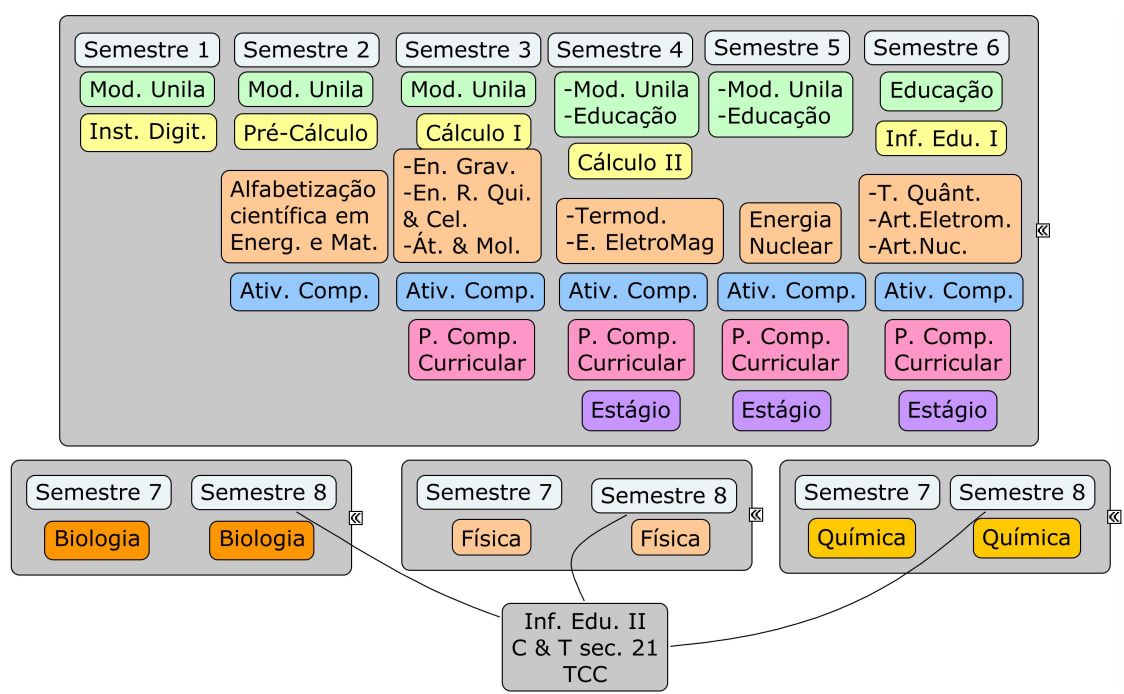

Figura 2 – Representação gráfica da estrutura curricular. As denominações dos componentes curriculares serão explicadas ao longo do texto.

A disciplina **Alfabetização Científica em Energia e Matéria** tem papel fundamental na estrutura pedagógica do curso. Em primeiro lugar, documentos oficiais e vários pesquisadores destacam a importância de textos de divulgação científica para o ensino básico, e por extensão para a formação de professores [44, 48, 51, 68-75]. Em segundo lugar, a disciplina cumpre um papel na definição de organizadores prévios ou conceitos ancoradouros, conforme a teoria da aprendizagem de Ausubel[7], na medida em que ao longo da disciplina os alunos terão a oportunidade de discutir temas atuais envolvendo os conceitos que serão trabalhados nas disciplinas de conteúdo científico na sequência do curso.

Os papéis desempenhados pelas três disciplinas sobre informática já foram mencionados acima.

A disciplina **Energia Gravitacional** (Energia G.) é basicamente aquilo que tradicionalmente se chama Física I, acrescida dos seguintes tópicos: formação da terra e o papel da gravidade na existência de nossa atmosfera; efeito da gravidade na estrutura e comportamento dos seres vivos; exploração espacial; usina hidroelétrica e outros artefatos mecânicos. Os tópicos acrescidos permitem conexão com as ciências da vida e com aplicações tecnológicas que se utilizam de conceitos da mecânica.

A disciplina **Energia nas Reações Químicas Celulares** (Energia Q. & Cel.) é uma introdução básica aos conceitos de bioquímica celular, tratando dos constituintes químicos da célula e derivação da energia, conversão de energia em forma biologicamente ativa, metabolismo celular, fotossíntese, síntese de proteínas e temas correlatos. O conteúdo é similar aos textos tradicionais, mas o tratamento diferencia pela ênfase que se dá ao conceito de energia.

Nas duas disciplinas anteriores, matéria e energia estão naturalmente presentes. Por exemplo, a energia gravitacional resulta de uma configuração de corpos massivos, sendo a energia potencial proporcional às massas, às distâncias entre os corpos, e ao campo gravitacional por eles produzidos. Na disciplina **Átomos e Moléculas** (Át. & Mol.) a matéria será inicialmente tratada isoladamente, a partir da tabela periódica e da teoria atômica. A tabela periódica oportuniza uma bela discussão sobre a estrutura eletrônica dos materiais e de suas propriedades físico-químicas. A disciplina finaliza com discussões elementares sobre processos de radiação, tais como corpo negro, efeito fotoelétrico, transições atômicas e nucleares.

**Termodinâmica** seria uma disciplina absolutamente idêntica àquela ministrada nas licenciaturas em física, se não fosse pelo acréscimo dos tópicos sobre energia e entropia em seres vivos e a discussão de aplicações tecnológicas. A grande novidade na presente proposta é que futuros professores de biologia e de química terão a oportunidade de apropriar-se dos conceitos da termodinâmica exatamente do mesmo modo como os futuros professores de física.

De modo análogo à disciplina Energia Gravitacional, **Energia Eletromagnética** (Energia EletroMag) é bastante similar à disciplina sobre eletricidade e magnetismo oferecida nos cursos de licenciatura em física, acrescida de tópicos pertinentes às ciências biológicas e da discussão qualitativa, para efeito de motivação, de contextos relevantes para o tema, como luz de uma vela, luz solar, efeitos biológicos das radiações, energia elétrica distribuída a partir de uma central hidroelétrica. Os temas pertinentes às ciências da vida são: o espectro eletromagnético e os seres vivos, sistemas de orientação e migrações, fototropismo, fototactismo e Biomagnetismo.

Como a estrutura CTS da presente proposta é baseada nos conceitos de energia e matéria como elementos transversais de toda a estrutura conceitual das ciências da natureza, e nos produtos resultantes das manipulações desses conceitos, ou seja os artefatos artesanais e industriais, é indispensável que alguns desses artefatos sejam discutidos no momento da correspondente apropriação conceitual, conforme veremos mais adiante. Englobamos as principais aplicações tecnológicas nos três tipos de artefatos abaixo descritos.

- ❖ **Artefatos Mecânicos**: Moinhos de roda. Sistemas de polias. Sistemas de catracas. Rotores. Turbinas hidráulicas. Motores de combustão interna.
- ❖ **Artefatos Eletromagnéticos**: Rádio. Telefone. Televisão. Radar. Forno de microondas. Dispositivos semicondutores. Laser. Produtos da Nanotecnologia.

- **Artefatos Nucleares**: Ressonância magnética nuclear. Radioterapia. Reatores para fusão nuclear. Aceleradores de partículas. Tomografia computadorizada. Tomografia por emissão de pósitrons.

## Discussão e Conclusões

Como é bem sabido, o desenvolvimento científico é um empreendimento humano que demanda trabalho cooperativo de um número incalculável de colaboradores, mediado por circunstâncias sócio-políticas e econômicas vitais para o seu desenvolvimento. Os produtos tecnológicos daí resultantes podem alterar profundamente essas circunstâncias sócio-políticas e econômicas, com efeitos eventualmente dramáticos nas relações internacionais. Portanto, o professor de ciências tem, como cidadão e formador de opinião, a obrigação de estar atento a essas circunstâncias da sua atividade profissional. A estrutura da presente proposta pedagógica facilita sobremaneira esse objetivo, como ilustrado a seguir.

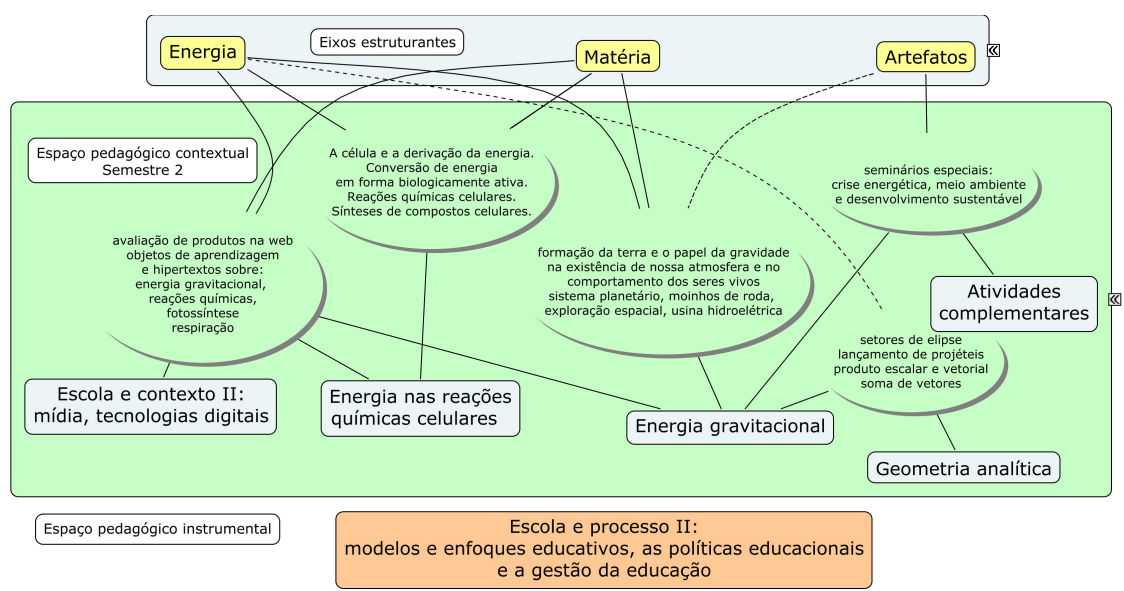

Figura 3 – Interconexões temáticas e conceituais no segundo semestre.

A disciplina Energia Gravitacional obrigatoriamente coloca em pauta as hidroelétricas. Portanto, é uma ótima oportunidade para se programar seminários, a título de atividades complementares, para discutir a crise energética, o impacto ambiental e o desenvolvimento sustentável.

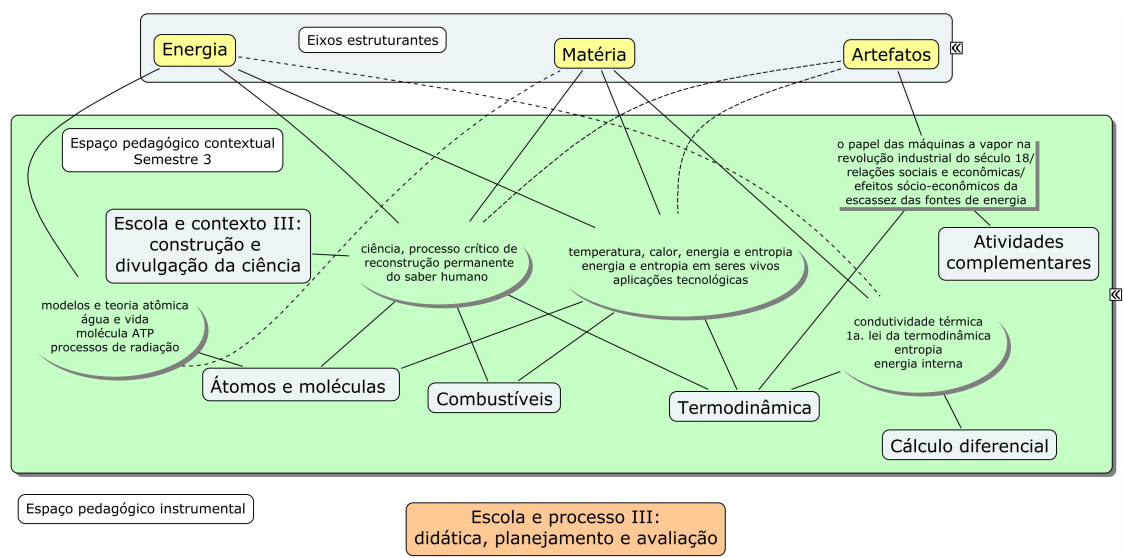

Figura 4 – Interconexões temáticas e conceituais no terceiro semestre.

É impossível estudar termodinâmica sem fazer referência à primeira revolução industrial, ocorrida no século 18, como também é impossível tratar daquela revolução industrial sem abordar as mudanças nas relações de trabalho. Portanto, no semestre em que essa disciplina é oferecida, atividades complementares podem ser programadas na forma de seminários, conduzidos por professores de história, geografia, economia, sociologia, entre outros.

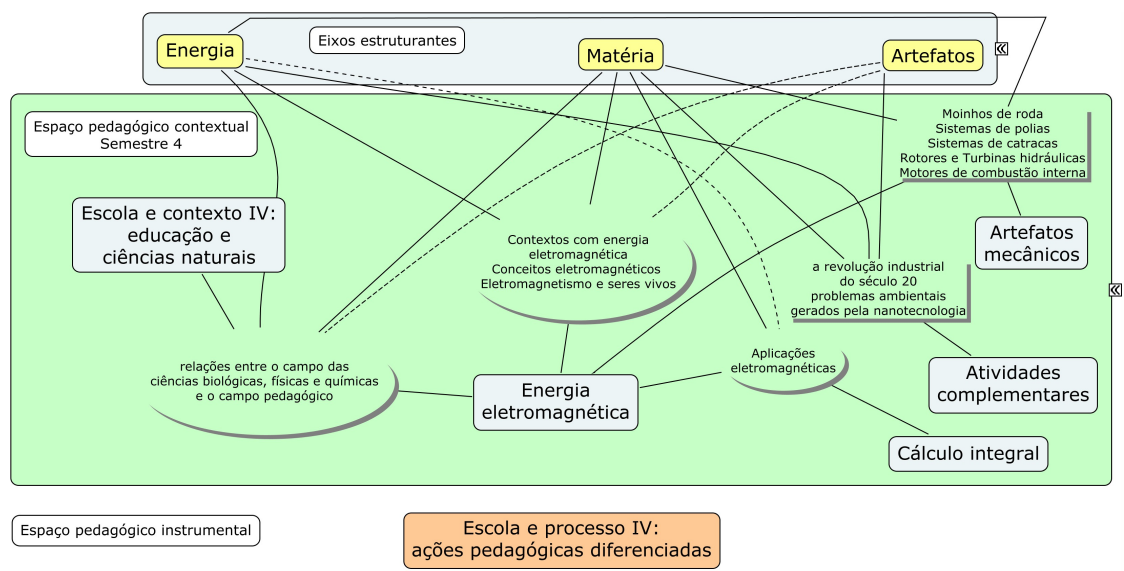

Figura 5 – Interconexões temáticas e conceituais no quarto semestre.

O quarto semestre é especialmente rico em termos de possibilidades de contextualizações sociais, políticas, culturais e econômicas. Nos semestres anteriores foram oferecidas as disciplinas Energia Gravitacional e Termodinâmica, de modo que no presente semestre, os alunos estudam a disciplina tecnológica associada, ou seja Artefatos Mecânicos. É interessante observar que dado o atual cenário tecnológico, com aplicações da microeletrônica e de nanotecnologia, essa disciplina funciona como uma espécie de curso de história dos primórdios da tecnologia.

No momento em que se estuda a Energia Eletromagnética, que ao lado da teoria quântica abre as portas para a tecnologia contemporânea, é bastante conveniente programar seminários sobre a revolução industrial do século 20 e o atualíssimo tema do impacto ambiental gerado pela nanotecnologia.

Para concluir, chamamos a atenção para o fato de que não discutimos as disciplinas ditas humanas (educação, sociologia, filosofia, história, entre outras), nem as de matemática, nem as específicas oferecidas no último ano, porque elas podem ser tratadas do mesmo modo como são nos atuais cursos de licenciaturas em biologia, física e química.

# Referências